\documentstyle[psfig]{l-aa}
\def\msol{{M}_{\odot}}
\begin{document}

   \thesaurus{03     
              ( 11.03.4; 
                11.09.3; 
		12.03.3; 
                12.04.1; 
		13.25.2)} 
%
%
   \title{ROSAT/HRI study of the optically rich, lensing cluster CL0500-24}

   \author{Sabine Schindler \inst{1,2} 
   \& 
   Joachim Wambsganss \inst{3} 
   }

   \offprints{Sabine Schindler}

   \institute{
	      MPI f\"ur extraterrestrische Physik,
	      Giessenbachstr.,
	      85748 Garching, 
	      Germany;
	      e-mail: {\tt sas@mpa-garching.mpg.de} 
	      \and
	      MPI f\"ur Astrophysik,
              Karl-Schwarzschild-Str. 1,
	      85748 Garching, 
	      Germany 
	      \and
	      Astrophysikalisches Institut Potsdam,
	      An der Sternwarte 16,
	      14482 Potsdam,
	      Germany;
	      e-mail: {\tt jwambsganss@aip.de}
	      }

   \date{}
   \maketitle

   \begin{abstract}
   An analysis of a ROSAT/HRI observation of the optically rich,
   gravitationally lensing galaxy cluster
   CL0500-24 (or Abell S0506) is presented. 
   We show that the X-ray luminosity of this supposedly rich cluster
   is relatively low at $1.1^{+0.2}_{-0.1}\times10^{44}$ erg/s  in the 
   ROSAT band, the bolometric X-ray luminosity is 
   $2.0^{+1.5}_{-0.7}\times 10^{44}$ erg/s.  The X-ray emission is
   strongly correlated with the northern subconcentration of the
   cluster galaxies at redshift $z = 0.327$.  
   The derived total mass of the (sub-)cluster within 1 Mpc 
   is $(1.5\pm 0.8) \times 10^{14}$M$_\odot$,
   with an upper limit to the gas mass of about $0.5\times 10^{14}$M$_\odot$,
   corresponding to a gas mass fraction of $\le 30^{+30}_{-10}$\%.
   The X-ray luminosity
   and the morphology of the extended X-ray emission 
   supports the view that CL0500-24 consists of two clusters at
   a velocity difference of $\Delta v \approx 3200$ km/s that
   happen to lie along the line of sight.

   \end{abstract}

%
%

\section{Introduction}

The cluster CL0500-24 is a medium redshift (z $\approx$ 0.32), 
rich cluster of galaxies (Giraud 1988). 
Optically there are two obvious subconcentrations  visible
(Infante et al. 1994). There is a straight blue arc  in the
cluster whose redshift ($z_{arc} = 0.913$) clearly identifies it 
as a background galaxy (Smail et al. 1993), thus confirming
the gravitational lens scenario suggested by Wambsganss et al. (1989). 

Since there has been some debate in the past whether the various
mass determinations for clusters of galaxies -- velocity
dispersions, gravitational lens masses, X-ray masses --
result in consistent or discrepant results (see e.g. Miralda-Escud\'e
\& Babul 1995; Tyson \& Fischer 1995; 
Smail et al. 1995; Squires et
al. 1996a, 1996b), this cluster seems to be a 
good target for another test of these methods against each other.

X-ray observations of a galaxy clusters can provide information
about the structure of clusters and therefore on their dynamical
state. This is of interest here, since the optical appearance
suggests physically distinct subclumps. 

Here we present X-ray observations with the ROSAT/HRI (Tr\"umper 1983).
We show the morphology of CL0500-24 as it appears in X-rays,
we determine an X-ray profile, and we estimate the mass of the
X-ray emitting gas and the total mass of the cluster. Finally, we
discuss our results and compare them with other X-ray/lensing clusters.
Throughout this paper we use $H_0=50$km/s/Mpc.

\section{ The cluster CL0500-24 (or Abell S0506) in optical light} 

The cluster CL0500-24 is a rich, very compact southern cluster
of galaxies (Giraud 1988) at  a redshift of $z \approx 0.32$
(in Abell et al. (1989) it is listed as supplementary cluster S0506).
There are 95-100 galaxies brighter than V = 23 within a radius of  
$\approx 0.55 h_{50}^{-1}$Mpc. The average density, N$_{0.5}$,
defined as  the number of bright galaxies,$ m \le  m_{3} + 2$ (where
$m_3$ is the magnitude of the third brightest galaxy)
projected within a radius of 0.5$h^{-1}_{50}$Mpc  is 31-36. This means
that CL0500-24 appears richer than the well studied giant
gravitational arc cluster Abell 370, for which $N_{0.5} = 28$,
see Mellier et al. (1988).
CL0500-24  contains a large fraction of blue 
galaxies, and has a quite high line of sight velocity dispersion of 
$\sigma = 1375 $km/s, based on redshifts of 22 cluster galaxies
(Giraud 1990). 
The cluster ``apparently has a triple core" (Giraud 1988)    so
it is to be expected that an X-ray map will  identify substructure.
Infante et al. (1994) also found a bimodal distribution in the
galaxy velocities, enhancing this view that the cluster consists
of subclumps.

The known redshifts  of the galaxies in CL0500-24 cover quite a broad 
range. 
Infante et al. (1994) showed that the velocity distribution 
(26 measured galaxies) is
bimodal with two peaks at $v_1  = 94817$ km/s   and 
                          $v_2  = 97987$ km/s. 
%
%
%
The velocity difference ($\Delta v = v_2 - v_1 = 3170$ km/s)
is much larger than the one-dimensional velocity dispersions of the two
sub-clumps ($\sigma_1 =  917\pm208$ km/s, 
            $\sigma_2 = 1152\pm214$ km/s).
This double nature   of the cluster was supported 
by  Infante et al. (1994)'s  group finding algorithms.
So it appears that CL0500-24 is not one rich cluster, but rather 
two clusters along the line of sight. 

The most striking feature in CL0500-24 is a straight, blue arc
with a length of about 14 arcseconds, which is about 22 arcseconds  away
from the apparent cluster center.
A gravitational lens model for this arc (Wambsganss et al. 1989)
finds very naturally a strongly elongated image of a 
background galaxy at the position  of the arc with the
correct orientation. This model predicts a mass of about
$1.4 \times 10^{14} $M$_{\odot}$ for the (projected) inner part of the
cluster, i.e. for a circle around the apparent center C
with radius center-arc (cf. Giraud 1990).
Equivalently, a cluster modeled as a singular isothermal sphere 
with a velocity dispersion of 1430 km/s 
is necessary to produce an arc with this length at this location.
The redshift of the arc was
recently measured to be $z_{arc} = 0.913$ (see Smail et al. 1993)
confirming the gravitational lens scenario quite satisfactorily.
So far no arclets or weak lensing signature
have been reported in this cluster.

\begin{figure*}
\psfig{figure=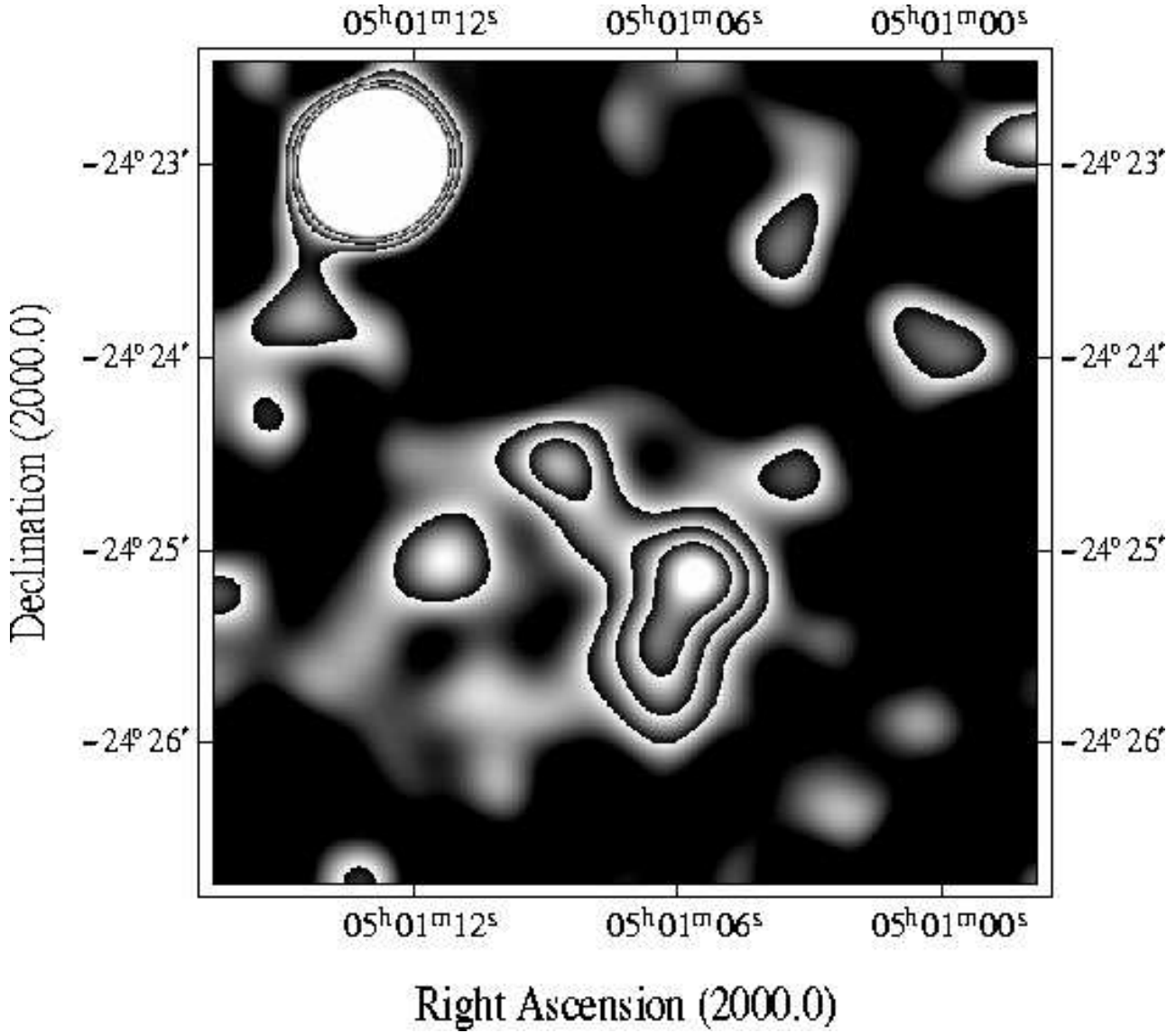,width=17.0cm,clip=} 
\caption[]{ROSAT/HRI image of the cluster CL0500-24.
It is smoothed with a Gaussian filter of $\sigma$ = 10 arcseconds.
The cluster has a clumpy structure with an extension south of the
maximum and additional emission in the north-east and in the east. The
bright X-ray source in the upper left corner is not associated with
the cluster.
}
\end{figure*}

\section{X-ray Data}

CL0500-24 was observed with the ROSAT/HRI in a pointed observation
with an exposure time of 37756 s between February $8^{th}$ and
13$^{th}$, 1995.

The positioning of the observation can be checked with a
point-like source which coincides with a star of blue magnitude 16.8
at a position $\alpha=05^h01^m26.0^s$, 
$\delta=-24^{\circ}17'31''$ (J2000). From this correspondence we infer
that the positional accuracy of the pointing is very good
(estimated error $\leq$ 4 arcseconds).
The total number of source photons for the cluster CL0500-24 
is 440. 

As the cluster emission is not very strong we  follow a
two-fold strategy. In addition to the straightforward analysis
of the full data set,  we try to improve the signal-to-noise ratio
by skipping the intervals with enhanced background. In practice,
we bin the data to intervals of 100 s. 
Subsequently, we drop the high background intervals with more 
than 500 counts / 100 s. The remaining time intervals with the
``cleaned data'' cover 57\% of the original exposure time.
The cleaned data set contains 270 source photons.

\subsection{Morphology}

Because of the limited number of photons from the
cluster the observed morphology is influenced by 
statistical fluctuations. 
In Fig. 1 we present an image smoothed with a Gaussian filter of
$\sigma=10$ arcseconds to give an idea how 
the morphology looks like. The main maximum of an image
smoothed with a Gaussian of $\sigma=5$ arcseconds is at
$\alpha=05^h01^m05.6^s$,  
$\delta=-24^{\circ}25'03''$ (J2000). 
It corresponds to a position between
the galaxies N and S (nomenclature according to Giraud (1990)) which are
identified by Infante et al. (1994) as the center of one of two
subconcentrations. In an image smoothed with a Gaussian of
$\sigma=10$ arcseconds the X-ray maximum is also very close to these
galaxies (see Fig. 2).
The X-ray emission is elongated southward
towards galaxy A. 
A second maximum of the X-ray emission can be seen 
in the North-East at $\alpha=05^h01^m09^s$, 
$\delta=-24^{\circ}24'31''$ (J2000), close to galaxy \# 12
(cf. Infante et al. 1994). There is no
extra emission associated with the second optical center C.
All these features are ``robust" in that they 
show up both in the ``cleaned'' and in the ``non-cleaned'' data.

The galaxies are assigned to two subclusters centred on galaxy N and on
galaxy C, respectively,  
by Infante et al. (1994) according to their velocities from
Giraud (1990) and Infante et al. (1994).
The distribution of N galaxies is much better correlated with the
X-ray emission than the distribution of C galaxies (see Fig. 2). 
This is a first
hint that we see in X-rays only (or mainly) the N subcluster.

In Fig. 3 the X-ray morphology is compared with the galaxy density 
(Infante, private communication) for
all galaxies, while in Fig. 2 only galaxies with measured velocities
are included. The luminosity weighted densities show again the two
main concentrations around N and C, out of which only N is correlated
with X-rays. Another maximum in the galaxy density in
the north-east is situated 
between two X-ray blobs and is probably belonging to  the N
concentration because there are three N galaxies in this region,
i.e. again no indication for any emission from the C subcluster.

\subsection{Luminosity}

The X-ray emission can be traced out to a radius of 2.5 arcminutes 
(860 kpc                     at distance of cluster).
As summarized in Table 1, we find a countrate of $0.012\pm 0.001$ 
counts/s within this radius.
To convert this countrate to a luminosity we assume a typical
metallicity of 0.35 solar (Arnaud et al. 1992; Ohashi 1995) and 
a Galactic
hydrogen column density of $7.12\times 10^{22}$ cm$^{-2}$ (Dickey \&
Lockman 1990). As there is no possibility to derive a gas temperature
with the HRI we assume a typical cluster temperature of 4 keV which is
also consistent with the $L_X-T$ relation (Edge \& Stewart 1991a;
David et al. 1993; White 1996).  With
these values the luminosity in the ROSAT band (0.1 - 2.4 keV) is
$1.1^{+0.2}_{-0.1}\times10^{44}$ erg/s and the bolometric luminosity
is $2.0^{+1.5}_{-0.7}\times 10^{44}$ erg/s. The errors include the 
uncertainties of
the countrate and an assumed
temperature range between 1 and 10 keV.

About 2.6 arcminutes north-east of the cluster centre is a point-like source
(see Fig. 1) at $\alpha=05^h01^m13^s$, 
$\delta=-24^{\circ}22'59''$ (J2000). It has a countrate of
$8.6\times10^{-3}$ counts/s. The position of this  source corresponds
within 10 arcseconds to the radio source PMN0501-2422 which is listed
in  Griffith et al. (1994) with a 4850 MHz flux density of 154 mJy.
Looking for an optical counterpart in the Southern Digitized Sky
Survey we find a faint object at about 4 arcseconds distance; it
cannot be distinguished whether it is a stellar object or a galaxy.

\begin{figure}
\psfig{figure=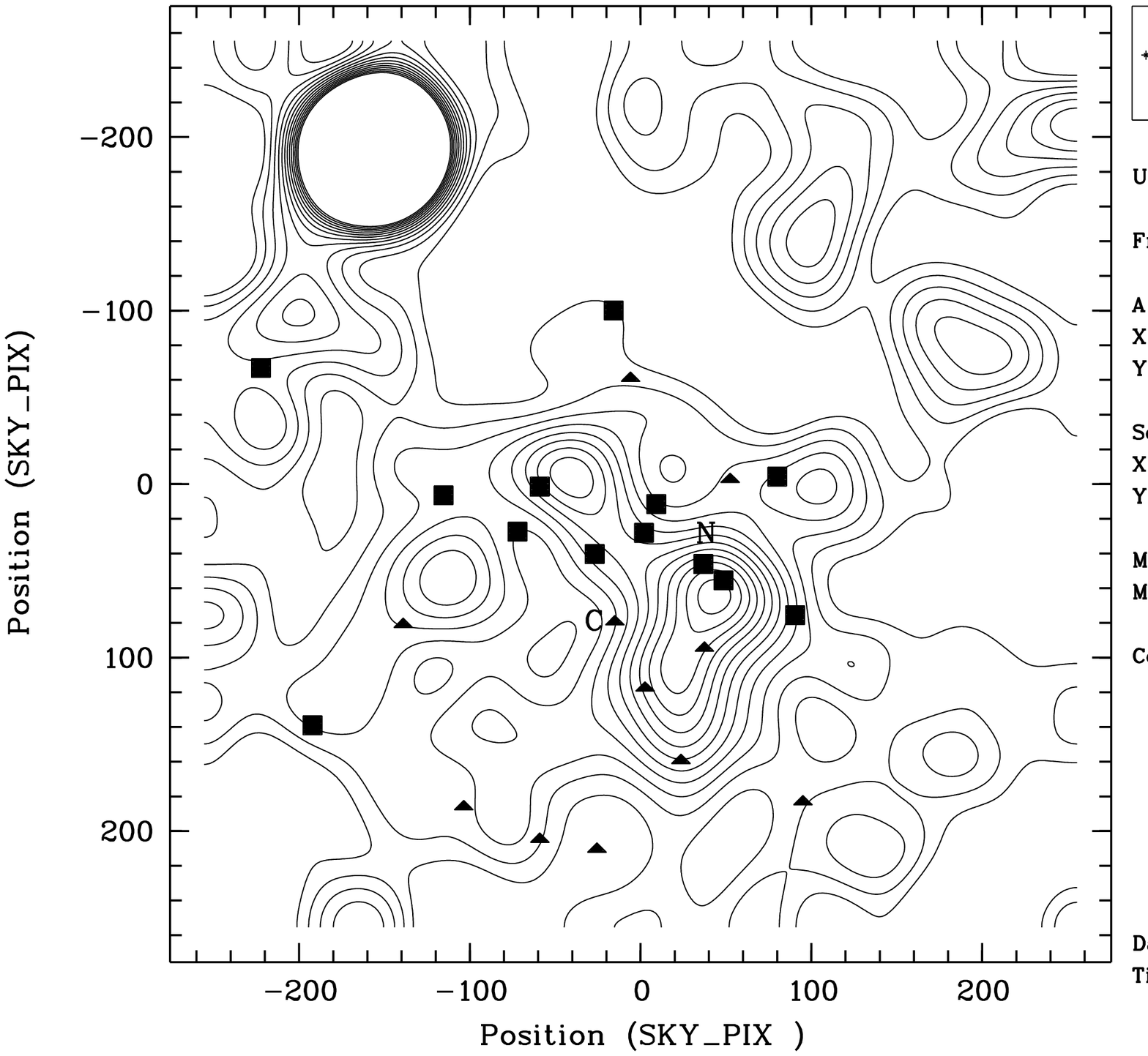,width=8.8cm,clip=} 
\caption[]{Cluster galaxies assigned to the subclusters N and C by
Infante et al. (1994) superimposed on the X-ray contours (smoothed
with a Gaussian filter of $\sigma$ = 10 arcseconds). Triangles:
galaxies assigned to subcluster C, squares: galaxies assigned to
subcluster N. The size of the image is the same as in Fig. 1. One
tickmark corresponds to 10 arcseconds. The
contour levels have a linear spacing of
$3.8\times10^{-4}$counts/s/arcmin$^2$. The highest contour corresponds
to $9.2\times10^{-3}$counts/s/arcmin$^2$, the lowest to 
$5.0\times10^{-3}$counts/s/arcmin$^2$. The central galaxies of the
subclusters are marked with N and C, respectively. The
X-ray emission is well correlated with the subcluster centred on N,
while there is hardly any correlation with the C subcluster. In
particular, there is no extra emission at the position of the central
galaxy C.
}
\end{figure}

\begin{figure}
\psfig{figure=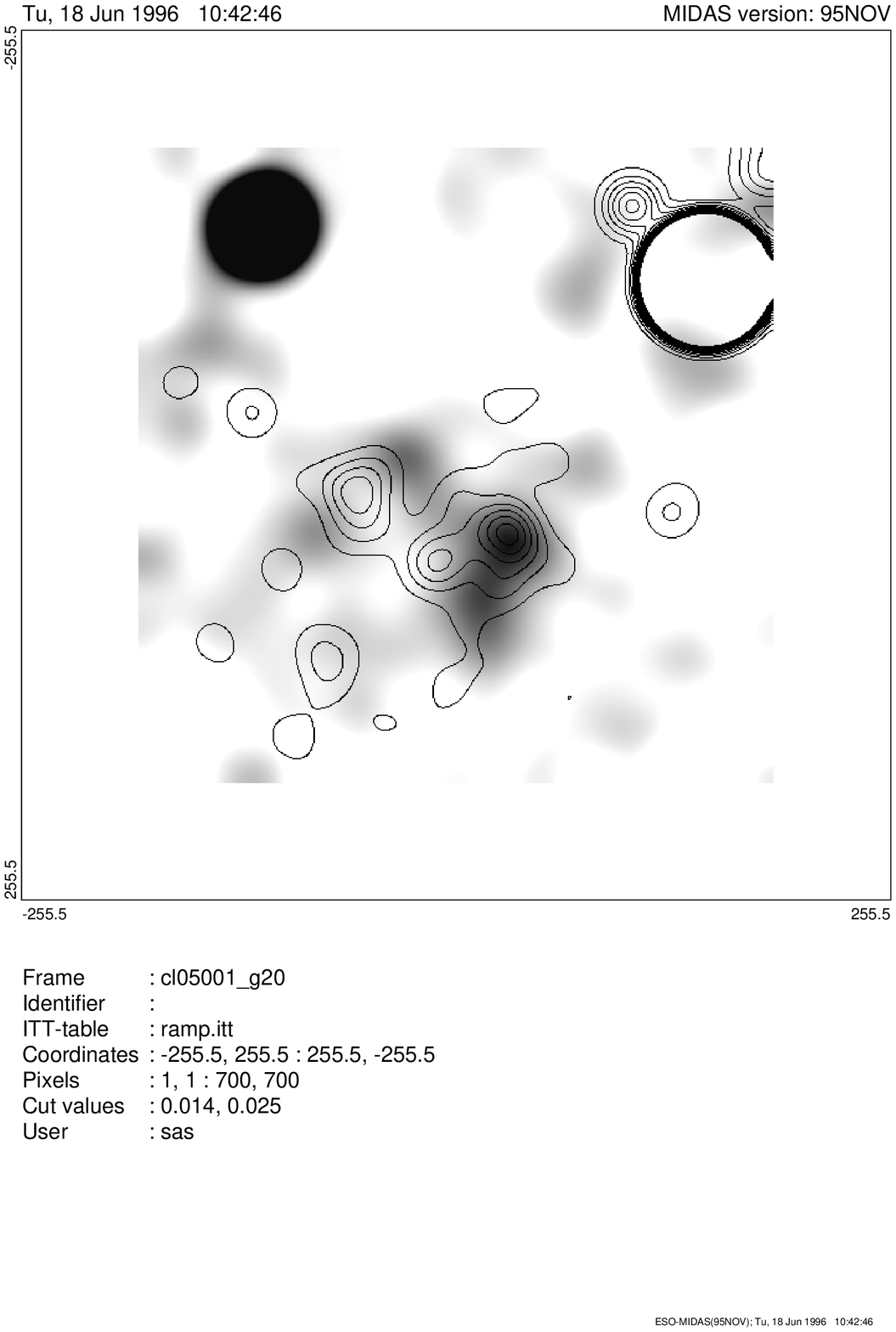,width=8.8cm,clip=} 
\caption[]{Luminosity weighted galaxy densities (contours) from V magnitudes
(Infante, private communication) smoothed with a Gaussian of
$\sigma=8$ arcseconds
are superimposed on the X-ray emission
(greyscales, same as the contours of Fig. 2). 
The size of the image is 2.5x2.5 arcminutes.
}
\end{figure}

\subsection{Profile}

Because of the limited number of photons and the non-spherical
appearance of the X-ray contours, 
the radial profile of the X-ray emission
is  not very well determined (Fig. 4). 
Nevertheless, we try to fit a $\beta$-model to
the surface brightness (following 
Cavaliere \& Fusco-Femiano 1976; Jones \& Forman 1984) 

$$\Sigma (r) = 
\Sigma_0 \left( 1 + {\left({r \over r_c}\right)}^2\right)^{-3\beta + 1/2},
   \eqno(1)
$$
where $\Sigma_0$ is the central surface brightness, $r_c$ is the core
radius, and $\beta$ is the slope parameter. The best fit values
averaged over a number of fits with or without ``cleaning'' and with
different binning centred on $\alpha=05^h01^m05.6^s$  
$\delta=-24^{\circ}25'03''$ (J2000) are  
$\Sigma_0= 8\times10^{-3}$ counts/s/arcmin$^2$, $r_c = 0.09$
arcminutes (30 kpc) and $\beta = 0.4$,                           a
very small $\beta$ and a very small core radius. 
But the 1$\sigma$
errors allow for a huge range of about 0.3 to 1.0 for $\beta$ and 
0 to 1.5 arcminutes (0 to 500 kpc) for the core radius.

\begin{figure}
\psfig{figure=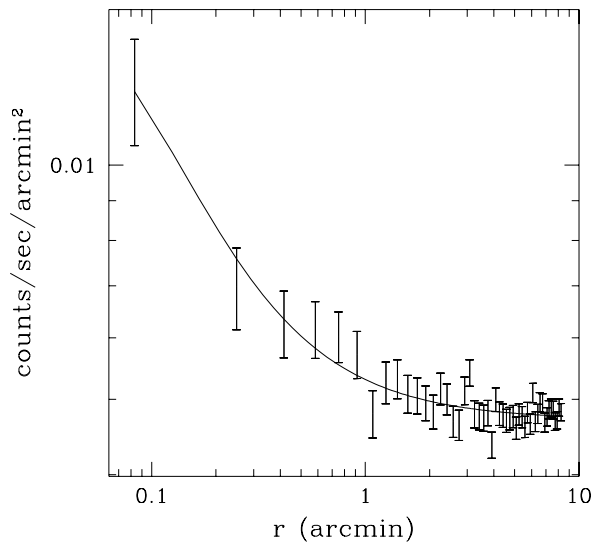,width=8.8cm,clip=} 
\caption[]{Radial profile of the X-ray emission fitted with a
$\beta$-model. 
}
\end{figure}

\subsection{Mass determination}

The parameters of the  $\beta$-model can be used to make
a deprojection of the  2D image 
to derive the three dimensional  density distribution. 
The profile of the integrated gas mass is shown in Fig. 5. 
Within a radius of  1 Mpc the gas mass amounts to
0.46$\times 10^{14}\msol$. 
As the emissivity of the gas  in the ROSAT energy band is almost
independent of the temperature (within the temperature range of 2-10 keV
it changes only by 6\%), we can derive the 
gas density distribution hardly affected by the uncertainty of the
temperature estimate. The only uncertainty are local unresolved 
inhomogeneities or substructure which result in an
overestimation of the gas mass. Therefore, the value given above is
strictly speaking an upper limit.

With the additional assumption of hydrostatic
equilibrium, the integrated total mass can be calculated from the 
equation
$$
M(r) = {-kr\over \mu m_p G} T \left({ d \ln \rho \over d \ln r }+
                                    { d \ln T    \over d \ln r }\right),
   \eqno(2)
$$
where $\rho$ and $T$ are the density and the temperature of the
intra-cluster gas, and $r$, $k$, $\mu$, $m_p$, and $G$ are the 
radius, the Boltzmann constant, the molecular weight, the proton mass, 
and
the gravitational constant, respectively. For the temperature we use
again a typical cluster temperature of 4 keV.
When using the parameters of the various $\beta$-model fits with
varying binning and with/out ``cleaning'' we find different
mass profiles only in the central part, as shown in the section
above. Beyond a radius of about 100 kpc the various mass profiles
are in good agreement. 
The error originating from the uncertainty in the temperature is 
certainly much larger and can amount to $\pm50\%$.

The results are shown in Fig. 5.
At a radius of 1 Mpc we find 
an integrated total mass of $(1.5\pm0.8)\times
10^{14}\msol$. The errors comprise a temperature range from 2 to 6
keV. This total mass yields a relatively high gas mass fraction
of $30_{-10}^{+30}$\%. The gas mass 
fraction is decreasing when going to smaller radii. But as the gas
mass fraction depends on $\beta$ and $\beta$ is not very well defined (see
Sect. 3.3) this decrease may not be very significant.

The total mass is even smaller than  the
luminosity-weighted virial mass 
of $2.2\times 10^{14}\msol$
of only the N concentration 
derived by Infante et al. (1994).
(The X-ray mass converted to their $H_0=60$km/s/Mpc would be
$1.3\times 10^{14}\msol$.)
This is
a hint that the X-ray emitting gas traces the potential of only
one subcluster.

For a comparison with mass estimates determined from the
gravitational lensing effect, we   
also calculate  the total mass 
of the cluster, as seen inside a certain angle, 
basically integrating a spherically symmetric
three dimensional mass distribution in cylindrical shells with 
cylinder axis parallel to the line of sight  
or integrating the cluster surface mass density $\Sigma(\theta)$ 
outward (see Fig. 5). 
The total mass inside a circle with radius cluster center-arc is
$0.3\times 10^{14}\msol$. From the point lens model, Wambsganss et
al. (1989) found a much larger value: $1.4\times 10^{14}\msol$
(or a velocity dispersion of $\sigma = 1197$ km/s for
an isothermal sphere model);
here we used   the now measured arc redshift of $z_{arc} = 0.913$ 
(Smail et al. 1993).
Obviously, the mass from the lens model is considerably higher than the 
mass obtained here when integrating outward the total mass  derived
from the X-ray emission.
But      the two mass estimates are not directly
comparable because the X-ray mass is centred on galaxy N while the lensing
model has an assumed centre close to `C'. 
There is a difference in these two mass determinations, 
but not a striking one.
However, this is not too surprising:  
the gravitational lens effect integrates all matter along the sight, and
we just pointed out that there are {\it two} clusters along the line
of sight.  
On the other hand, 
the X-ray mass determination is most sensitive to the highest density
regions in the center of the cluster.
Furthermore, the point lens or spherically symmetric isothermal
sphere model 
used for the gravitational lens model (Wambsganss et al. 1989)
is certainly too simple to account in detail for the action of two 
subconcentrations along the line of sight.


\begin{figure}
\psfig{figure=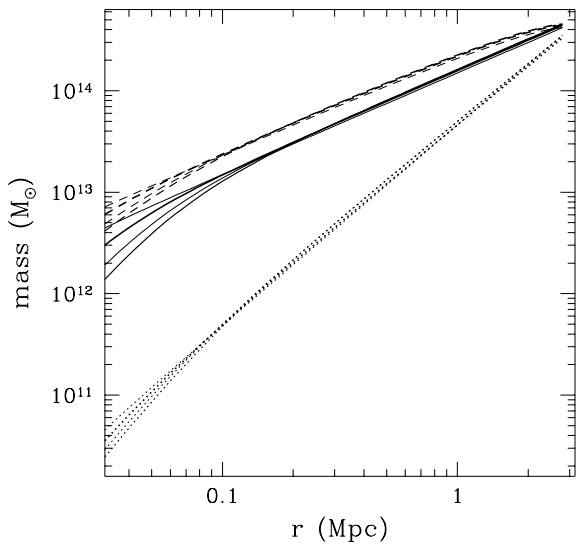,width=8.8cm,clip=} 
\caption[]{Mass profile of Cl0500-24. The dotted lines show the
integrated gas mass, the solid lines the integrated total
mass. Profiles for data with/out ``cleaning'' and with
different binning are plotted. In the inner region the scatter is
quite large because of the limited number of photons. But beyond a
radius of 100 kpc the profiles agree well.  For comparison
with mass determinations by the gravitational lens
effect, the dashed lines depict the integrated
surface mass density.

}
\end{figure}

\begin{table}[htbp]
\begin{center}
\begin{tabular}{|l|l|}    
	\hline
	&\\
	countrate(0.1-2.4keV) & $0.012\pm0.001$ counts/s \\
	&\\
	$L_X$(0.1-2.4keV) & $1.1^{+0.2}_{-0.1}\times10^{44}$erg/s \\
	&\\
	$L_X$(bol)       & $2.0^{+1.5}_{-0.7}\times10^{44}$erg/s \\
	&\\
	$M_{gas}$($r<1$Mpc)     & $0.5\times 10^{14}\msol$\\
	&\\
	$M_{tot}$($r<1$Mpc)     & $1.5\pm0.8\times 10^{14}\msol$\\
	&\\
	gas mass fraction ($<1$Mpc) & $30^{+30}_{-10}$\% \\         
	&\\ 
	\hline 
\end{tabular}
\end{center}
\caption{ Summary of the X-ray properties of CL0500-24}
\end{table}

\section{Discussion and conclusions}

The X-ray luminosity of CL0500-24 is quite moderate. 
Actually,
it is very small for a ``rich" cluster of galaxies. 
The Abell Cluster A370 (at a comparable redshift),
which  supposedly is about  equally rich 
as CL0500-24, is roughly ten times more luminous  in X-rays
(Lea \& Henry 1988; Fabricant et al. 1991; Bautz et al. 1994). 
Does this show that there is a wide spread of X-ray luminosities
for clusters that appear similarly rich optically? 

We argued that the X-ray emission of CL0500-24 originates
mainly from the northern subclump. 
But even when comparing the bolometric luminosity of CL0500-24
only with the velocity dispersion of the Northern subcluster, 
it is unexpectedly low (Edge \& Stewart 1991b).
This   indicates at least a moderate spread
in the optical richness versus X-ray luminosity relation.

There are other examples that illustrate this,  e.g. the
two distant clusters CL0939+4713 and CL0016+16, which are both
optically very rich (Dressler 1994; Koo 1981) but have quite different
bolometric X-ray luminosities: $1.1\times 10^{45}$erg/s (Schindler \&
Wambsganss 1996) and $5.6\times 10^{45}$ erg/s (Neumann \& B\"ohringer
1996), respectively.

Given the bimodal velocity distribution of the galaxies in 
CL0500-24 (cf. Chapter 2), there are two possibilities for the
spatial distance of the subclusters in the line of sight. One
possible scenario is that the two subclusters are relatively unrelated and
the different 
velocities are caused mainly by the Hubble flow. 
In this case they have   a distance of more than 60 Mpc. 
The other possibility is that the two subclusters have
a physical distance that is smaller than that corresponding to
the redshift difference, and are in the process of
colliding. 
As the relative velocity of almost 3200 km/s is very high,
in such a scenario
the subclusters would have to pass right through each other 
along the line of sight. In fact, the projected distributions of the
galaxies with redshift $z_2$ preferentially north, and with $z_1$
preferentially south indicates, that such a collision could not 
be exactly along the line of sight, but with some angle relative
to it. 
So the velocity 
difference is only the contribution projected onto the line of sight,
and the real 3dim relative velocity must be even larger than the 
measured difference.
N-body models show that relative velocities around 3000 km/s or more
are reached only during a very short period of time as the subclusters
pass through each other (Schindler \& B\"ohringer 1993; Huss et
al. 1996). 
If the subclusters are just passing through each other, an 
X-ray emission enhanced by a about a factor of two relative
to a quiescent state is expected during a
collision (Schindler \& M\"uller 1993). 
However, 
given the low observed X-ray luminosity, and the fact that it is
very unlikely that we see CL0500-24 exactly in this extreme short 
phase corresponding to such an encounter with high relative 
velocity we conclude that it is extremely unlikely that the
two subclumps are currently undergoing such a collision.

This view of two unrelated clusters 
is supported both by the morphology and the luminosity
of the X-ray emission of the cluster. 
The X-rays originate
from regions that are related to the subclump around N with
galaxy velocities peaking at $v_2$. 
There is no significant detection from the supposed center of
the cluster near the galaxy  C (cf. Giraud 1990) or other
galaxies at redshift $z_1$. 

The mass assigned by Infante et al. (1994) 
to the   more massive subclump around N
at $z_2$ is even higher than the X-ray mass
determined here. That is an indication that the 
X-rays trace the potential of only one subcluster while the
southern subcluster around galaxy C apparently 
does not affect the X-ray mass determination.
The mass found by Wambsganss et al. (1989) for the cluster from
a simple gravitational lens model for the straight arc is 
higher than the X-ray mass determined here. This also
fits into this picture qualitatively, 
because the gravitational lens effect
integrates all the mass along the line of sight, and so unavoidably
considers all the subclumps discussed here.

%
%


The total mass of ($1.5\pm0.8)\times 10^{14}\msol$ within 1 Mpc which is
about the virial radius according to the spherical collapse
approximation (Gunn \& Gott 1972;
White et al. 1993) is a relatively low value compared to the typical
mass range of clusters of $5-50\times10^{14}\msol$ (B\"ohringer
1995). It is an order of magnitude lower than the mass of the nearby
clusters Coma and Perseus (B\"ohringer 1994). The gas mass fraction of
30\% is at the upper limit of the typical range for clusters of
10-30\% (B\"ohringer 1995).

Combined, all the arguments mentioned above point into the direction
that CL0500-24 consists of two concentrations with little
direct interaction. 
Only one of the subclusters has X-ray emission. 
But even considering the fact that only part of the optically visible
cluster is associated with the X-ray properties,
the values for the X-luminosity and the derived total mass are
comparably small.


%
%

\begin{acknowledgements}
It is a pleasure to thank Hans B\"ohringer, Makoto Hattori, Andreas
Huss, and Doris Neumann for helpful discussions. 
We also thank Leopoldo Infante for providing to us unpublished galaxy data. 
S.S. acknowledges 
financial support by the Verbundforschung.

\end{acknowledgements}
%
%

\end{document}